\documentclass[aps,prd,reprint]{revtex4-1}

\usepackage{graphicx}

\usepackage{amsmath}

\usepackage{placeins}

\usepackage{xspace}

\begin{document}

\title{Nonlinear Saturation of Kinetic Ballooning Modes by Zonal fields in Toroidal Plasmas}
\author{G. Dong,$^1$ J. Bao,$^2$ A. Bhattacharjee,$^1$ and Z. Lin$^2$ }
\affiliation{$^1$Princeton Plasma Physics Laboratory, Princeton University, New Jersey 08540}
\affiliation{$^2$Department of Physics and Astronomy, University of California, Irvine, California 92697, USA}
\date{\today}

\begin{abstract}
Kinetic ballooning modes (KBM) are widely believed to play a critical role in disruptive dynamics as well as turbulent transport in tokamaks. While the nonlinear evolution of ballooning modes has been proposed as a mechanism for ``detonation" in tokamak plasmas, the role of kinetic effects in such nonlinear dynamics remains largely unexplored. In this work global gyrokinetic simulation results of KBM nonlinear behavior are presented. Instead of the finite-time singularity predicted by ideal MHD theory, the kinetic instability is shown to develop into an intermediate nonlinear regime of exponential growth, followed by a nonlinear saturation regulated by spontaneously generated zonal fields. In the intermediate nonlinear regime, rapid growth of localized current sheet is observed. 
\end{abstract}
\maketitle

Ballooning instability (or its astrophysical counterpart, the Parker instability) in a magnetized plasma is characterized by long wavelength parallel and short wavelength perpendicular to the background magnetic field. The nonlinear evolution of the instability has been a subject of great interest for a diverse range of eruptive phenomena in space plasmas such as substorms in the Earth's magnetotail \cite{Hurricane1999,zhu2007} and edge-localized modes (ELMs) \cite{kirk2006} and disruption precursor modes \cite{tftr} in toroidal fusion plasmas. Theoretical studies of nonlinear ideal magnetohydrodynamic (MHD) ballooning modes \cite{cowley1997, wilson2004} predict explosive nonlinear growth near the linear instability threshold. Finger-like structures develop, forming a front with steep pressure gradient which can further destabilize the mode nonlinearly, and result in a finite-time singularity (``detonation") \cite{cowley1997}. However, attempts at simulating such an instability using the full MHD equations have not succeeded in realizing the tendency to form a finite-time singularity. While finger-like structures and the steepening of pressure gradients are indeed observed \cite{zhuprl} as predicted by the analytical theory, the mode is seen to grow in the nonlinear regime exponentially with approximately its linear growth rate, maintaining its filamentary structure. A new asymptotic regime, called the ``intermediate" nonlinear regime of exponential growth has been formulated analytically to account for these simulations \cite{zhu2009}. 

	Exponential nonlinear growth, even though it lacks the impulsive signature of ``detonation", is of great interest. During this phase, the width of the finger-like structures becomes sufficiently small that the validity of the MHD model is questionable. In weakly collisional plasmas, kinetic effects will intervene. This leads to considerations of the kinetic ballooning mode (KBM) which is widely recognized to play an important role in fusion \cite{kirk2007, zhu2009} as well as space plasmas \cite{pritchett2009}. However, the nonlinear dynamics of the KBM, and in particular the role of kinetic effects \cite{tang1980} is not well understood. Numerical simulations on the nonlinear saturation of KBMs have been inconclusive \cite{ishizawa15, waltz10, ma2017}. Various simulation results suggest that KBMs can only saturate nonlinearly with pressure profile relaxation \cite{ma2017} or increased flow shear \cite{waltz10} above a critical value of beta.

	In this Letter, we demonstrate from a gyrokinetic particle-in-cell (PIC) global simulation that following a linear regime of exponential growth, the KBM evolves into an intermediate regime which exhibits features that are qualitatively different from its ideal MHD counterpart \cite{zhuprl}. First, the growth is slightly faster than linear exponential growth, which indicates that the perfect cancellation that occurs in the intermediate ideal MHD dynamics between nonlinear destabilization due to enhanced pressure gradients and stabilization due to field-line bending \cite{zhu2009} does not occur in the kinetic intermediate regime. Second, and more important than the first, the kinetic electromagnetic dynamics lead to the spontaneous generation of zonal flow (flux-surface-averaged electrostatic potential $\left\langle \delta\phi\right\rangle$) and zonal current (flux-surface-averaged vector potential $\left\langle\delta A_\parallel\right\rangle$) through three-wave coupling processes, in contrast to the generation of zonal flow through modulational instability in electrostatic ion-temperature-gradient (ITG) turbulence \cite{lchen}. When the zonal flow shear exceeds the linear growth rate, zonal flow shearing suppresses the nonlinear instability which in turn, self-regulates the zonal fields (the zonal flow and the zonal current), leading to a saturated nonlinear regime. In the kinetic intermediate regime, thin current sheets develop near the mode rational surfaces, which can eventually exhibit tearing instability in the presence of resistivity, but the resistive taring mode growth rate appears to be too slow to have a strong effect on the nonlinear dynamics of the instability, which is dominated by the regulation of the mode by zonal fields.

\textit{\textbf{Gyrokinetic simulation of KBM.}}---
In the simulations using the gyrokinetic toroidal code (GTC) \cite{linscience, holod15}, ions are treated by the gyrokinetic Vlasov equation, while electrons are described using the nonlinear fluid equations: the electron perturbed density $\delta n_e$ is calculated by time-advancing the continuity equation \cite{wenjun12}, and the electron parallel flow $\delta u_{\parallel e}$ is calculated by inverting the parallel Ampere's Law \cite{holod09}. The gyrokinetic Poisson's equation is solved to obtain the perturbed electrostatic potential $\delta \phi$. The parallel vector potential $\delta A_\parallel=\delta A_\parallel^{adi}+\delta A_\parallel^{na}$ is solved for the adiabatic and non-adiabatic parts. Integrating the electron drift kinetic equation to the momentum order and considering the parallel Ampere's law, we can derive the linear Ohm's law for adiabatic $\delta A_\parallel^{adi}$ and the nonlinear Ohm's law for non-adiabatic $\delta A_\parallel^{na}$ as:    
\begin{equation}
\frac{\partial \delta A_\parallel^{adi}}{\partial t}=\frac{c}{B_0}\mathbf{B_0}\cdot\nabla \delta\phi_{ind}
\end{equation}
and
\begin{equation}
\begin{split}
&\frac{1}{c}\frac{\partial \delta A_\parallel^{na} }{\partial t} = \frac{\mathbf{\delta B}_\perp}{B_0}\cdot\nabla\delta\phi_{ind} \\
&-\frac{m_e}{n_0e^2}\nabla\cdot\left(\delta u_{\parallel e}\frac{cP_{e0}\mathbf{B_0}\times\nabla\delta B_\parallel}{B_0^3}\right)+\frac{P_{e0}}{en_0}\frac{\mathbf{\delta B_\perp}}{B_0^2}\cdot\nabla\delta B_\parallel,
\end{split}
\end{equation}
where $\mathbf{B_0}$ is the equilibrium magnetic field, $\mathbf{\delta B_\perp}=\mathbf{\delta B}_\perp^{adi}+\mathbf{\delta B}_\perp^{na}$ is the perturbed perpendicular magnetic field, which consists of the adiabatic and non-adiabatic parts as $\mathbf{\delta B}_\perp^{adi}=\nabla\delta A_\parallel^{adi}\times\mathbf{B_0}/B_0$, and $\mathbf{\delta B}_\perp^{na}=\nabla\delta A_\parallel^{na}\times\mathbf{B_0}/B_0$, and $\delta B_\parallel$ is the compressional magnetic perturbation, which is solved using the gyrokinetic perpendicular force balance equation \cite{dong}. Here $\delta\phi_{ind}=\frac{T_e}{e}\left(\frac{\delta n_e}{n_0}-\frac{\delta\psi^{adi}}{n_0}\frac{\partial n_0}{\partial \psi_0}\right)-\delta\phi$ is the inductive potential. $\delta \psi^{adi}$ is the adiabatic component of the perturbed poloidal flux, which is defined as $\nabla\delta\psi^{adi}\times\nabla\alpha=\mathbf{\delta B}_\perp^{adi}$, where $\alpha=q(\psi_0)\theta-\zeta$ is the fieldline label with the Boozer poloidal angle $\theta$ and toroidal angle $\zeta$, and the safety factor $q(\psi_0)$ is a function of the equilibrium poloidal flux $\psi_0$. Also, $T_e$ is the electron equilibrium temperature, $n_0$ is the plasma equilibrium density, and $P_{e0}=n_0T_e$ is the electron equilibrium pressure. The first term on the right-hand-side of Eq. (2) represents the nonlinear ponderomotive force, which is shown in the following sections to play an important role in KBM nonlinear dynamics. The second and third terms are the nonlinear drive from finite $\delta B_\parallel$, and are small compared with the nonlinear ponderomotive drive due to the smallness of $\beta$. A complete form of the generalized Ohm's law is presented in \cite{bao2017}. In addition to the zonal current, the nonlinear ponderomotive force in Eq. (2) can produce localized current sheets on the mode rational surfaces with tearing parity. Since the typical current sheet width realized in our simulations is larger than the electron skin depth, we drop the electron inertia terms in the generalized Ohm's law. The equilibrium current $u_{\parallel 0}=-\frac{c}{4\pi en_{0}B_0}\mathbf{B_0}\cdot\nabla\times \mathbf{B_0}$ is included in the simulation. The flux-surface-averaged component of the Poisson's equation and Eq. (2) are solved for the zonal flow and the zonal current respectively. 

In the simulations, Cyclone Base Case parameters are used for the background plasmas: the major radius is $R_0=83.5 cm$, inverse aspect ratio is $a/R_0=0.357$. At $r=0.5a$, the plasma parameters are $B_0=2.01T$, $T_e=2223eV$, $R_0/L_T=6.9$, $R_0/L_n=2.2$, $q=1.4$, $\beta_e=2\%$. The first order $s-\alpha$ model with circular cross-section \cite{wenjun12} is used for the equilibrium magnetic field, with $\theta=\theta_0-\epsilon\sin\theta_0+O(\epsilon^2)$, where $\theta_0$ is the geometric poloidal angle, and $\epsilon=r/R_0$ is the normalized radial coordinate. We simulate n=10 toroidal mode KBM (keeping all the poloidal harmonics $m$), and its nonlinear interaction with the zonal mode (m=0, n=0). In the linear simulations, the mode exhibits ballooning mode structures, with a real frequency $\omega_r^{lin}=0.67 c_s/a$ and maximum growth rate $\gamma^{lin}=0.62 c_s/a$. Convergence studies show that the physical results in the linear and nonlinear simulations are not sensitive to grid size, time step size or number of particles per cell.

\begin{figure}[]
\includegraphics[width=0.47\textwidth]{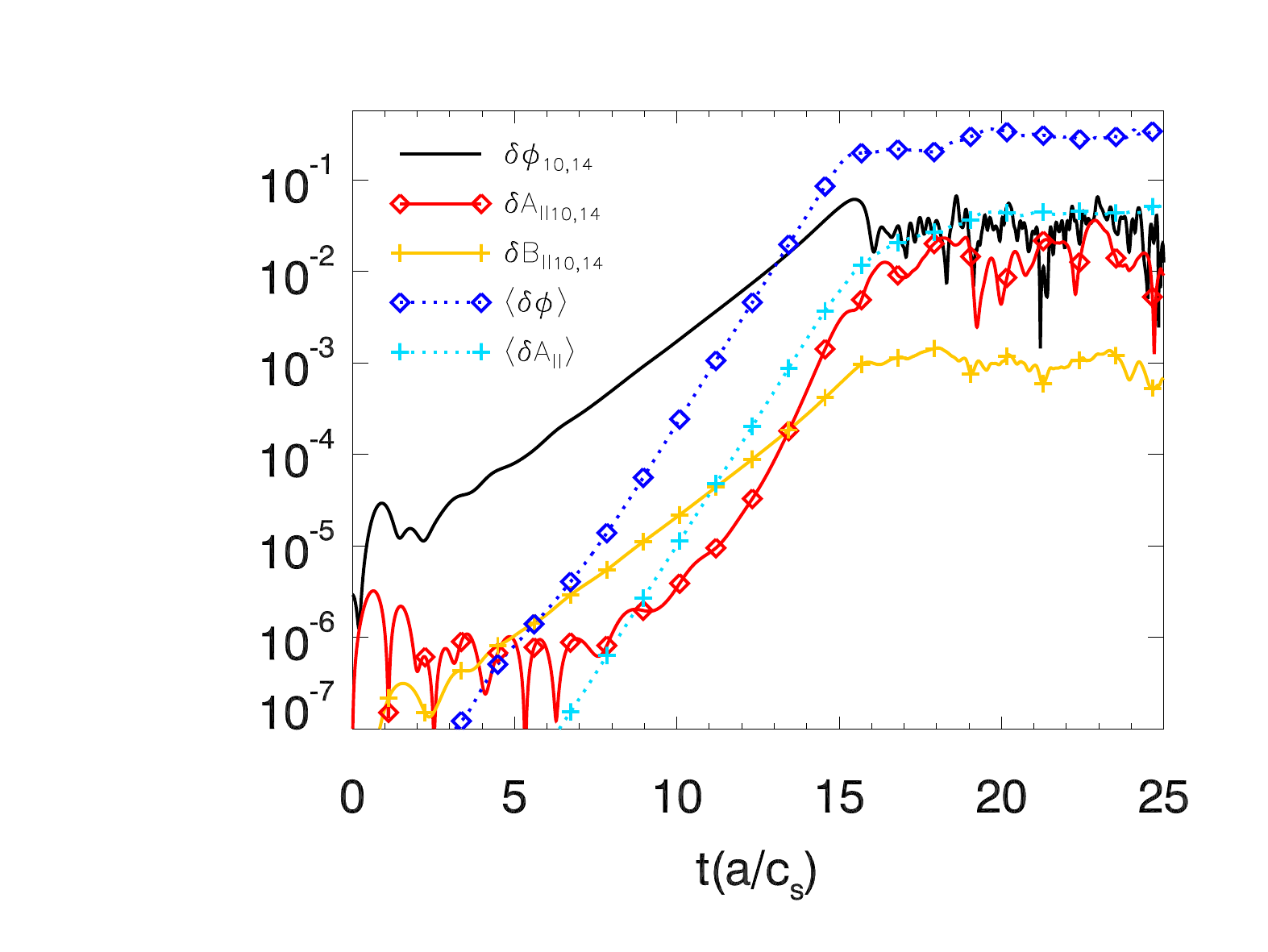}
\caption{Time history of the normalized perturbed electrostatic potential $\delta\phi$, parallel vector potential $\delta A_\parallel$, parallel magnetic field $\delta B_\parallel$ for mode $(m=14, n=10)$ measured at the $(14, 10)$ rational surface, and radial averaged zonal flow and zonal current amplitude.}
\end{figure}

\textbf{\textit{Saturation by zonal fields.}}---
A time history for the nonlinear KBM simulation is shown in Figure 1. The perturbed electrostatic potential, parallel vector potential, and parallel magnetic field are normalized, consistent with ideal MHD scaling, as $e\delta\phi/T_e, c\delta A_\parallel/v_A B_0 R_0$ and $\delta B_\parallel/B_0$ respectively, where $v_A=B_0/\sqrt{4\pi n_0 m_i}$. Normalization will be omitted in the following text for simplicity. The perturbed electrostatic potential $\delta\phi_{10,14}$, the parallel vector potential $\delta A_{\parallel 10, 14}$  , and the parallel magnetic field $\delta B_{\parallel 10,14}$ of the dominant $(10, 14)$ mode are measured at the mode rational surface with $q=1.4$ at the center of the simulation domain, and the zonal flow $\left\langle\delta \phi\right\rangle$ and the zonal current $\left\langle\delta A_{\parallel}\right\rangle$ amplitude are averaged over the simulation domain. The mode starts as a small perturbation in the fields, and quickly evolves into the linear regime after a brief transient stage. Before $t\sim 11 a/c_s$, $\delta\phi_{10,14}$ grew more than two orders of magnitudes at the linear growth rate $\gamma^{lin}$. Here $\delta A_{\parallel 10, 14}$ remains around three orders of magnitude lower than $\delta\phi_{10,14}$, since the linear adiabatic component $\delta A_{\parallel 10,14}^{adi}$ is zero at the resonant surface, as constrained by Eq. (1). Also, $\delta B_\parallel$ is more than one order of magnitude smaller than $\delta\phi$ due to the small plasma $\beta$.  At $t\sim 11 a/c_s$, $\delta A_{\parallel 10, 14}$ starts to grow faster than exponential, indicating that the mode evolves into an intermediate regime, where the nonlinear ponderomotive effects become important. From $t\sim 11 a/c_s$ to $t\sim 15 a/c_s$, $\delta \phi_{10,14}$ and $\delta B_{\parallel 10,14}$ grow slightly faster than exponential, with an effective growth rate $\gamma^{int}= 1.1\gamma^{lin}$ between $t=11 a/c_s$ and $t=14.5 a/c_s$, and the field quantities retain their linear poloidal mode structure. These features of KBM in this intermediate regime are similar to those in the intermediate regime found in compressible MHD simulations \cite{zhuprl}. The growth of dominant field quantities at a rate faster than predicted by linear theory indicates that exact cancellation of nonlinearities in the ideal MHD intermediate regime does not occur in this kinetic intermediate regime \cite{zhu2009}. In the linear regime and the intermediate regime, the zonal flow and the zonal current both grow exponentially at a growth rate $\gamma^{zonal}\sim 2\gamma^{lin}$. This shows that the zonal fields in KBM are driven more efficiently by the three-wave parametric instability, in contrast to the zonal flow excitation by modulational instabilities in electrostatic ITG, where the zonal flow grows exponential on exponential and the instantaneous growth rate depends on mode amplitude \cite{lchen}.

\begin{figure}[]
\includegraphics[width=0.47\textwidth]{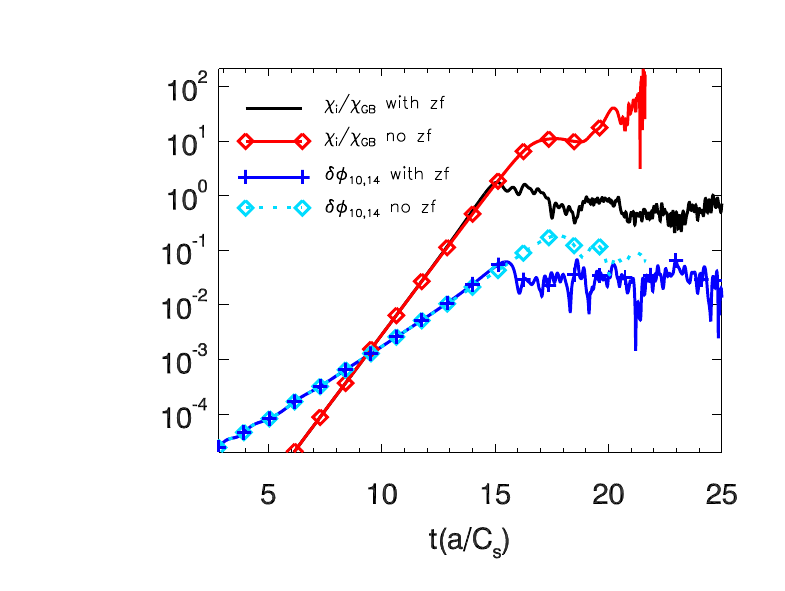}
\caption{Time history is shown for the ion heat conductivity $\chi_i$ and the perturbed electrostatic potential $\delta\phi$ for mode $(m=14, n=10)$ measured at the $(14,10)$ rational surface in a simulation with self-consistently generated zonal flow and zonal current and in a simulation with artificially suppressed zonal fields.}
\end{figure}

At $t\sim 15 a/c_s$, the dominant mode and the zonal fields saturate nonlinearly. As shown by the diamond dotted blue line in Figure 1, the steady state zonal flow amplitude is around $5$ times larger than the dominant $\delta\phi_{10,14}$ component. At the same time, the ion energy transport reaches steady state at the gyro-Bohm level with $\chi_i\sim\chi_{GB}$, as shown by the black solid line in Figure 2, where $\chi_{GB}=\rho_i^2v_i/a$, $v_i=\sqrt{T_i/m_i}$, and $\rho_i=v_im_ic/eB_0$ . The ion heat conductivity $\chi_i=\frac{1}{n_0\nabla T_i}\int d\mathbf{v} (\frac{1}{2}m_iv^2-\frac{3}{2}T_i)v_{r}\delta f$ is defined as the volume averaged ion energy flux normalized by local temperature gradient, where $v_r$ is the radial drift velocity including the $E\times B$ drift and the magnetic flutter drift \cite{bao2017}. In the simulations where the zonal flow and the zonal current are both artificially suppressed, the nonlinear ion heat conductivity becomes one order of magnitude larger, as shown by the diamond red line in Figure 2. Note that $\delta\phi_{10,14}$ also saturates at a magnitude around 3 times higher than that in the case with both zonal flow and zonal current, as shown by the diamond dotted blue line in Figure 2. In two other simulations where only the zonal current or the zonal flow  is artificially suppressed, $\delta\phi$ and $\chi_i$ saturation levels also see a significant increase, indicating that both zonal flow and zonal current regulate ion energy transport and mode saturation in KBM. A comparison of $\delta\phi$ nonlinear poloidal structure between simulations with and without the zonal fields is shown in Figure 3. In the simulation with self-consistently generated zonal flow and zonal current, the zonal fields break up the radially elongated eigenmode structure into micro and mesoscale structures, reducing radial transport. In the simulation with the zonal fields artificially suppressed, although the non-zonal nonlinear $E\times B$ term also shears the mode structure, some macroscale radial filaments of streamers survive. These results show that the KBM saturation is governed by the zonal fields, including both the zonal flow and the zonal current. In two additional simulations where $\beta_e=1.74\%$ and $\beta_e=1.55\%$ (near the KBM instability threshold), we observe similar nonlinear saturation features.

\begin{figure}
\includegraphics[width=0.23\textwidth,trim={{0.24\textwidth} {0.75\textwidth} {0.4\textwidth} {0.1\textwidth}},clip]{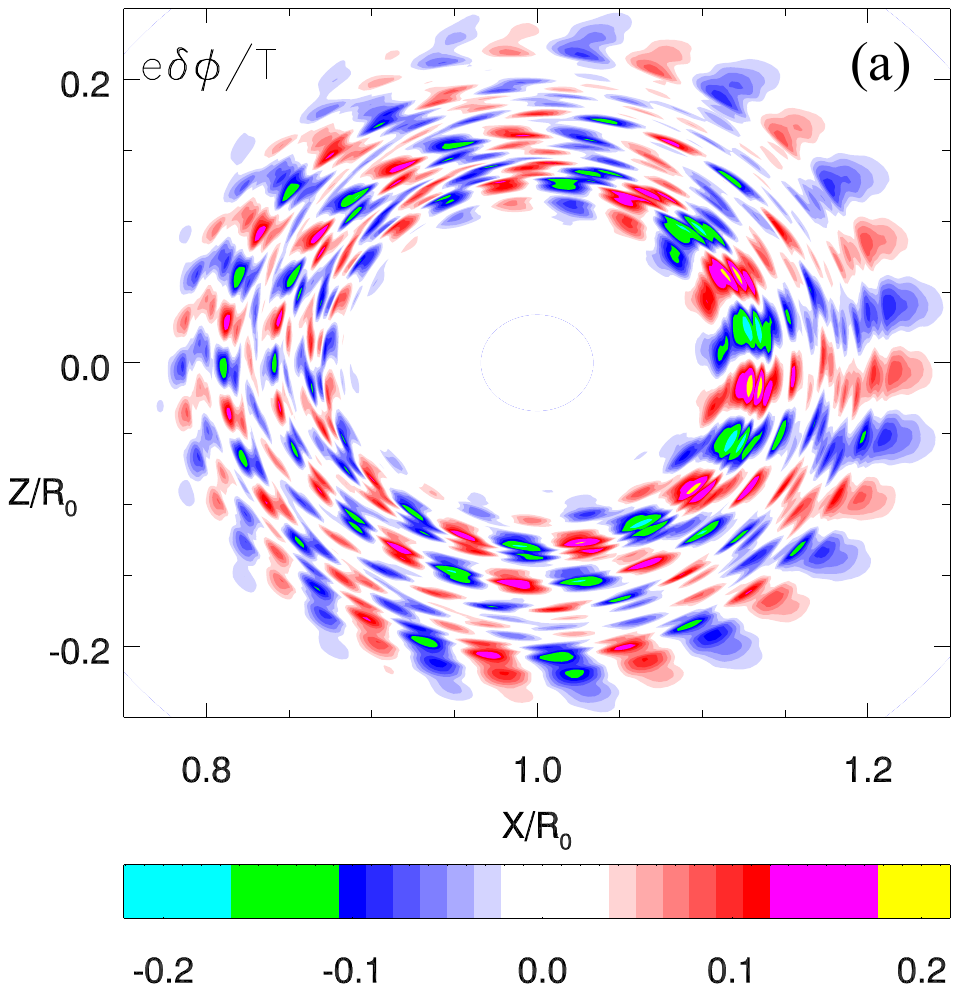}
\includegraphics[width=0.23\textwidth,trim={{0.24\textwidth} {0.75\textwidth} {0.4\textwidth} {0.1\textwidth}},clip]{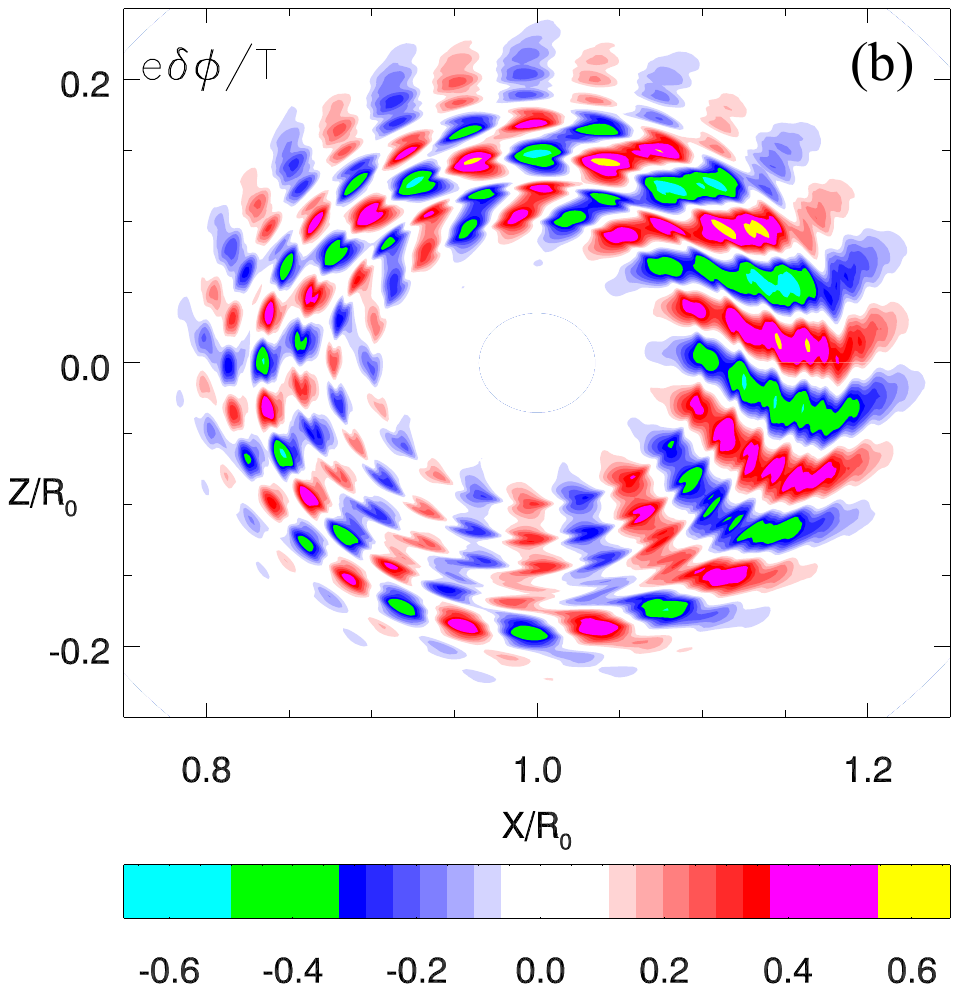}
\caption{Poloidal contour of the perturbed electrostatic potential $\delta \phi$ at the nonlinear regime. The left panel (a) shows broken radial filaments in the simulation with self-consistently generated zonal flow and zonal current. The right panel (b) shows macroscale radial filaments in the simulation with the zonal fields artificially suppressed. To clearly illustrate the difference in radial filaments, the $\left\langle\delta \phi\right\rangle$ component is not plotted in (a).}
\end{figure}

\textbf{\textit{Onset of nonlinear rapid growth of localized current sheet.}}---
As shown by the diamond solid red line in Figure 1, $\delta A_{\parallel 14,10}$ at the mode rational surface first grows faster than exponential, and then grows more than one order of magnitude exponentially with a nonlinear growth rate $\gamma^{nl}\sim 3 \gamma^{lin}$ during the intermediate nonlinear regime. The nonlinear growth rate can be explained by the coupling between the zonal current and non-zonal inductive potential through the first term in Eq. (2). The poloidal $\delta A_\parallel$ structure evolves from the linear eigenmode structure at $t=11 c_s/a$, as shown in Figure 4(a), to mesoscale structures at $t=17 c_s/a$, as shown in Figure 4(b). The mode structure becomes very thin in the radial direction, and remains the same scale as the linear eigenmode structure in the poloidal direction. This corresponds to the rapid growth of current sheets localized at the rational surfaces, excited by the nonlinear ponderomotive force terms in the generalized Ohm's law. In the simulation where the nonlinear ponderomotive force terms are not included ($\delta A_\parallel^{na}=0$), although zonal flows still break the linear mode into mesoscale structures nearly isotropic in radial and poloidal directions, as shown in Figure 4(c), the radial correlation length of the turbulence eddies is much longer than that in the case with the self-consistent ponderomotive force. 

The development of localized current sheet in the intermediate and nonlinear regime in KBM is analogous to the nonlinear process in the ideal MHD theory. However in this scenario where the kinetic effects become important during the intermediate regime, the mode saturates at the spatial scale comparable to the ion gyroradius with a transport level controlled by the zonal fields. In contrast, the mode structure in the MHD theory tends to become singular until the pressure profile flattens by transport. The radial profiles of (n,m) harmonic of $\delta A_\parallel$ at $t=11 c_s/a$ and $t=17 c_s/a$ are shown in Figure 4(d) and Figure 4(e). The linear mode structure has exact odd parity at the rational surfaces, and the nonlinear mode structure contains even parity component at the rational surfaces driven by the nonlinear electromagnetic ponderomotive force. For comparison, Figure 4(f) shows the (n,m) harmonic of $\delta A_\parallel$ after saturation in the simulation with $\delta A_\parallel^{na}=0$. In this case, although the mode structures deviate from the linear mode structures, each (n,m) harmonic is still zero at the $q=m/n$ surface. Finally, the mode exhibits a forward cascade in the radial mode number $k_r$ as it evolves to the nonlinear regime, corresponding to the nonlinear shearing and breaking of mode structures as shown in the poloidal contours. Because of the formation of thin current layer near rational surfaces, we conducted simulations with finite resistivity in the generalized Ohm's law to test the role of resistive tearing physics in the saturation of KBM \cite{dliu14}. With resistivity $100$ times the Spitzer resistivity, no significant tearing instability is observed within the time scale of KBM nonlinear saturation. In this case, KBM linear growth rate and real frequency are increased significantly by the resistive drive, and the zonal fields still saturate the mode with radially smoother nonlinear mode structure.

\begin{figure*}[]
\includegraphics[width=0.31\textwidth,trim={{0.24\textwidth} {0.75\textwidth} {0.4\textwidth} {0.1\textwidth}},clip]{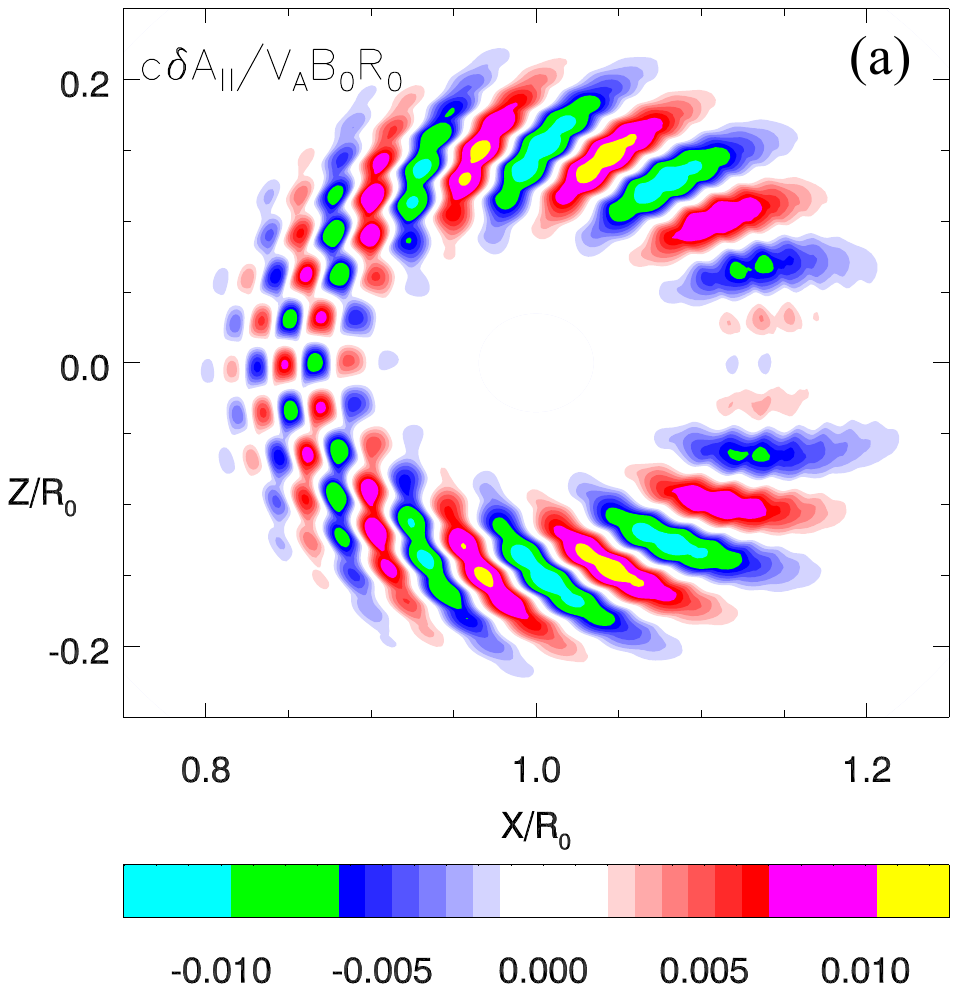}
\includegraphics[width=0.31\textwidth,trim={{0.24\textwidth} {0.75\textwidth} {0.4\textwidth} {0.1\textwidth}},clip]{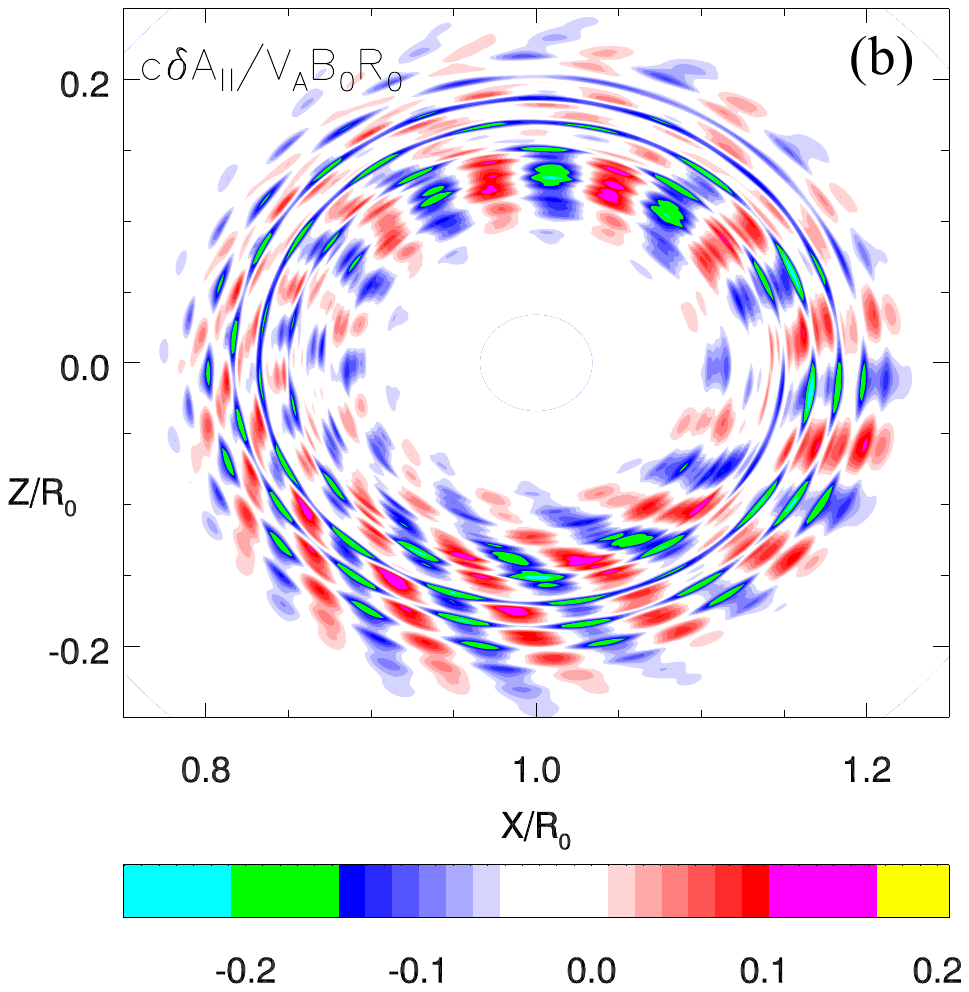}
\includegraphics[width=0.31\textwidth,trim={{0.24\textwidth} {0.75\textwidth} {0.4\textwidth} {0.1\textwidth}},clip]{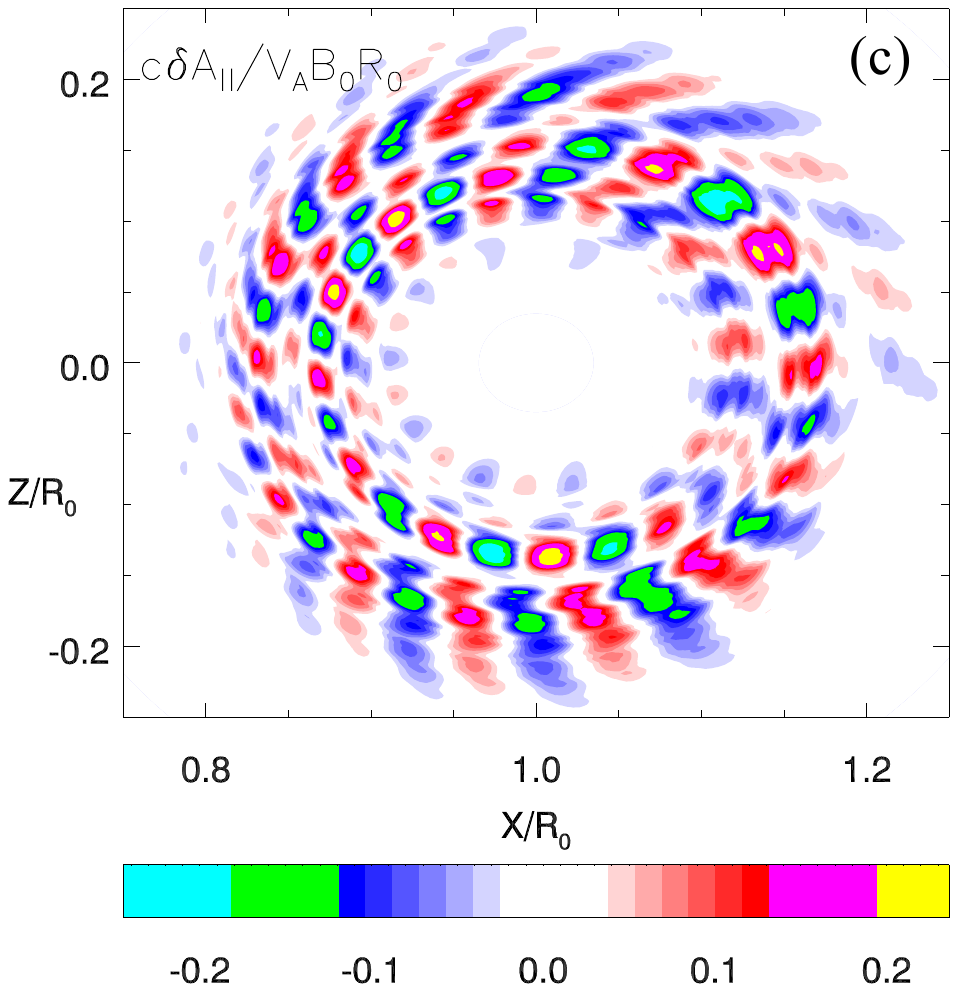}
\includegraphics[width=0.3\textwidth,trim={{0.25\textwidth} {1.0\textwidth} {0.493\textwidth} {0.18\textwidth}},clip]{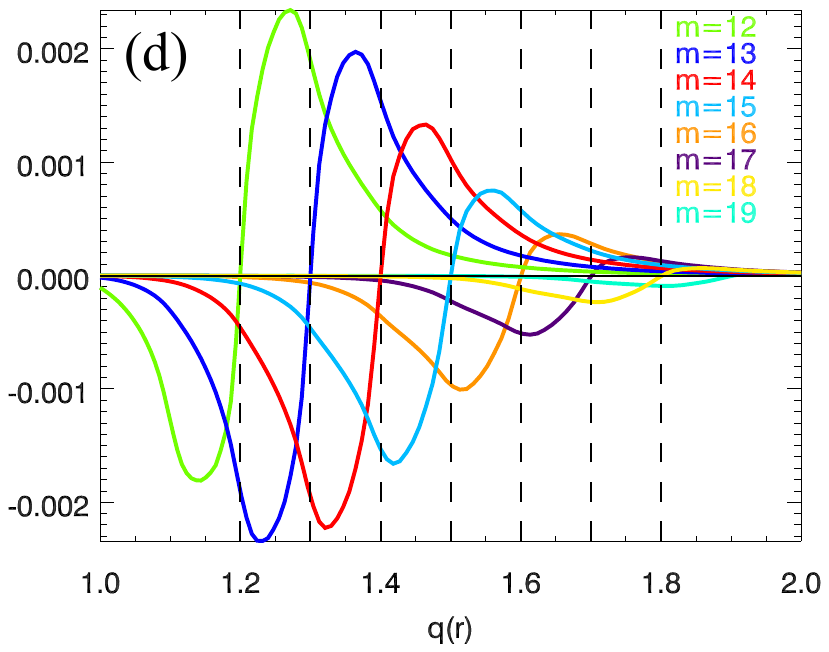}
\includegraphics[width=0.3\textwidth,trim={{0.25\textwidth} {1.0\textwidth} {0.493\textwidth} {0.18\textwidth}},clip]{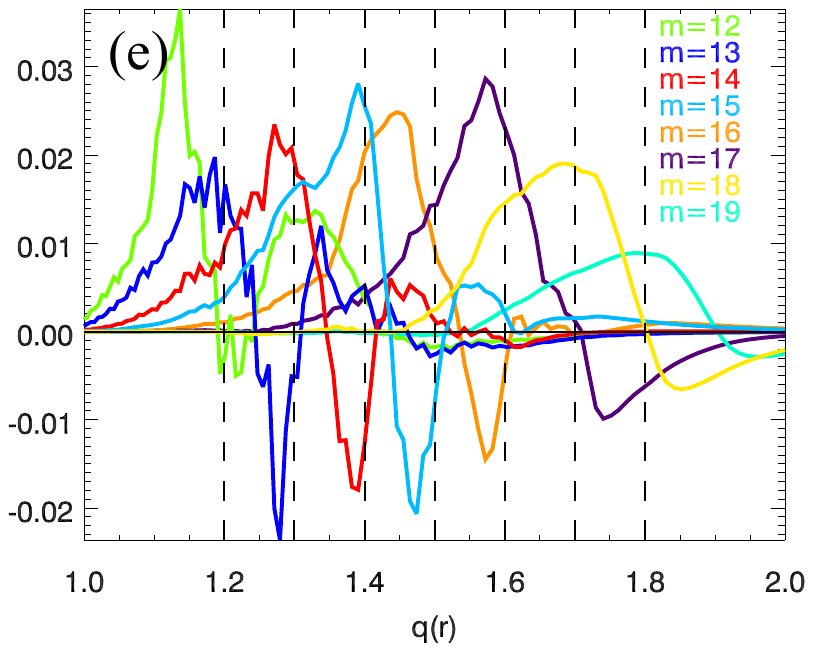}
\includegraphics[width=0.3\textwidth,trim={{0.25\textwidth} {1.0\textwidth} {0.493\textwidth} {0.18\textwidth}},clip]{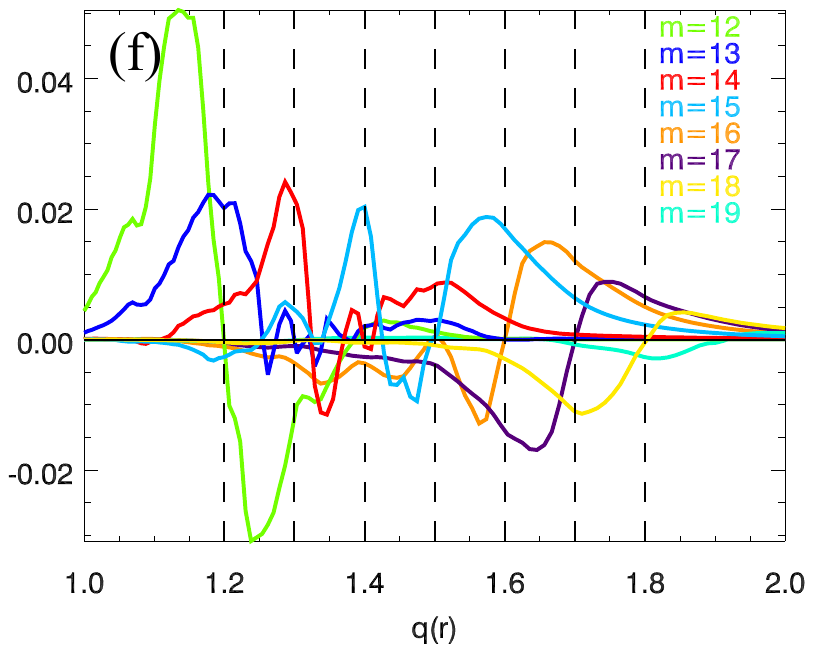}

\caption{Poloidal contour of the parallel vector potential $\delta A_\parallel$ linear structure before the intermediate regime in the top left panel (a), and $\delta A_\parallel$ nonlinear structure after the intermediate regime in the top middle panel (b) in the simulation with self-consistent ponderomotive force. Top right panel (c) show poloidal contour of nonlinear $\delta A_\parallel$ in the simulation without the ponderomotive force terms but with zonal flows. Corresponding lower panels (d), (e) and (f) show the radial profile of $(n,m)$ harmonic of $\delta A_\parallel$ in (a), (b) and (c).}
\end{figure*}

\textit{\textbf{Conclusions and future work.}}---
In summary, we have presented global gyrokinetic particle-in-cell simulation results of KBM nonlinear behavior. Instead of finite-time singularity, the instability develops into an intermediate nonlinear regime of exponential growth, followed by nonlinear saturation regulated by spontaneously generated zonal fields. Both zonal flow and zonal current are important for the saturation level. In the intermediate nonlinear regime, onset of rapid nonlinear growth of localized current sheet is observed. Although we did not observe significant growth of resistive tearing mode on the time scale of KBM nonlinear evolution, the current sheet near the rational surfaces might induce tearing instabilities on a longer time scale, which can affect the transport level, and provide seed islands for global tearing modes such as the neoclassical tearing mode. Therefore including the collisionless tearing physics by keeping the electron inertia terms in the simulation model \cite{bao2017} is an important next step. On the other hand, KBM turbulence can induce large transport at plasma edge in H-mode plasmas \cite{cmod,d3d,asdex}. According to the EPED model \cite{snyder2009}, KBM and peeling ballooning mode (which has both ballooning drive and bootstrap current drive) together determines the pedestal height and width. In future work, we plan to extend the KBM simulations presented in this Letter to experimentally measured equilibrium profiles with multiple modes to determine the prediction of KBM turbulence transport, its coupling with tearing modes and its role in shaping the edge.

In this work, research is supported by U.S. Department of Energy grants DE-AC02-09CH11466, DE-FG02-07ER54916, (DOE) SciDAC ISEP Center. This work used resources of the Oak Ridge Leadership Computing Facility at the Oak Ridge National Laboratory (DOE Contract No. DE-AC05-00OR22725) and the National Energy Research Scientific Computing Center (DOE Contract No. DE-AC02-05CH11231).

\end{document}